\documentstyle[12pt]{article}
\newcommand{\beq}{\begin{equation}}
\newcommand{\eeq}{\end{equation}}
\newcommand{\beqa}{\begin{eqnarray}}
\newcommand{\eeqa}{\end{eqnarray}}
\newcommand{\ba}{\begin{array}}
\newcommand{\ea}{\end{array}}
\begin{document}

$\;\;\;$
\vskip 1.2 truecm
\begin{center}
{\bf COEXISTENCE OF ORDERED AND CHAOTIC STATES \\IN NUCLEAR STRUCTURE}\\
\vspace{1.cm}
{V.R. Manfredi$^{1,2}$, \underbar{L. Salasnich}$^{2,3}$} \\
\vspace{1.2cm}
{$^1$ Dep. of Physics, Univ. of Padova, Italy, 
$^2$ INFN, Padova, Italy,\\
$^3$ Dep. of Pure and Appl. Math., Univ. of Padova, Italy} 
\end{center}

\vspace{1.4cm}

\section{Introduction}
In atomic nuclei, as in other many--body systems, the classical phase 
space is mixed, so ordered and chaotic states generally coexist \cite{first}. 
In this contribution we discuss some models, showing the transition from 
order to chaos. In several cases a clear correspondence between classical 
and quantum chaos has been established. 
In particular the transition from ordered to chaotic states will be 
discussed in the framework of the shell and roto--vibrational models. 
The spectral statistics of 
low--lying states of several $fp$ shell nuclei are studied with realistic 
shell--model calculations \cite{second}. 
Furthermore, for the roto--vibrational model the Gaussian curvature 
criterion of the potential energy 
clearly shows the transition from order to chaos for different 
values of rotational frequency \cite{third}. 

\section{Shell Model Calculations}

In this section we discuss the statistical analysis of the shell--model 
energy levels in the $A=46$--$50$ region. 
By using second--quantization notation, 
the nuclear shell--model Hamiltonian can be written as 
\beq
H=\sum_{\alpha} \epsilon_{\alpha}a_{\alpha}^+a_{\alpha} 
+\sum_{\alpha \beta \gamma \delta} <\alpha \beta|V|\delta \gamma >
a_{\alpha}^+ a_{\beta}^+ a_{\gamma} a_{\delta} \; ,
\eeq 
where the labels denote the accessible single--particle states, 
$\epsilon_{\alpha}$ is the corresponding single--particle energy, 
and $<\alpha \beta|V|\delta \gamma >$ is the two--body matrix element 
of the nuclear residual interaction. 
\par
Exact calculations 
are performed in the ($f_{7/2}$,$p_{3/2}$,$f_{5/2}$,$p_{1/2}$) 
shell--model space, with $^{40}$Ca as an inert core, 
by using a fast implementation of the Lanczos algorithm with 
the code ANTOINE. The interaction we use is a minimally 
modified Kuo--Brown realistic force with monopole improvements \cite{second}. 
\par
The spectral statistic $P(s)$ may be used 
to study the local fluctuations of the energy levels. 
$P(s)$ is the distribution of nearest--neighbour spacings 
$s_i={\tilde E}_{i+1}-{\tilde E}_i$ of the unfolded levels. 
To quantify the chaoticity of $P(s)$ in terms of a parameter, 
it can be compared to the Brody distribution, 
\beq
P(s,\omega)=\alpha (\omega +1) s^{\omega} \exp{(-\alpha s^{\omega+1})} \; ,
\eeq
with $\alpha = (\Gamma [{\omega +2\over \omega+1}])^{\omega +1}$. 
This distribution interpolates between the Poisson distribution ($\omega =0$) 
of regular systems and the Wigner distribution 
($\omega =1$) of chaotic ones. 
The parameter $\omega$ can be used as a simple 
quantitative measure of the degree of chaoticity. 
\par
The following table shows the calculated Brody parameter $\omega$ 
for the nearest neighbour level spacings distribution 
for $0\leq J\leq 9$, $T=T_z$ states up to $4$, $5$ and $6$ MeV above the 
yrast line in the analyzed nuclei.

\begin{center}
\begin{tabular}{|ccccccccc|} \hline\hline 
Energy & $^{46}$V & $^{46}$Ti & $^{46}$Sc & $^{46}$Ca & $^{48}$Ca & 
$^{50}$Ca & $^{46}$V+$^{46}$Ti+$^{46}$Sc & $^{46}$Ca+$^{48}$Ca+$^{50}$Ca \\ 
\hline
$\leq 4$ MeV & 1.14 & 0.90 & 0.81 & 0.41 & 0.58 & 0.67 & 0.92 & 0.56\\ 
$\leq 5$ MeV & 1.10 & 0.81 & 0.96 & 0.53 & 0.58 & 0.69 & 0.93 & 0.60\\ 
$\leq 6$ MeV & 0.93 & 0.94 & 0.99 & 0.51 & 0.66 & 0.62 & 0.95 & 0.61\\ 
\hline\hline
\end{tabular}
\end{center}

\par
We observe that Ca isotopes are less chaotic than their neighbours. 
In fact, the two--body matrix elements of the 
proton--neutron interaction are, on average, larger than those of 
the proton--proton and neutron--neutron interactions. 
Consequently, the single--particle mean--field motion in nuclei with 
only neutrons in the valence orbits suffers 
less disturbance and is thus more regular. 

\section{The Roto--Vibrational Model}

The Hamiltonian of an axially symmetric roto--vibrational nucleus 
is given by \cite{third}
\beq 
H={1\over 2}B(3a_0^2+2a_2^2)\omega^2 + 
{1\over 2}B({\dot a}_0^2+2{\dot a}_2^2) + V(a_0,a_2) \; ,
\eeq
with
\beq
V(a_0,a_2)= {1\over 2}C_2(a_0^2+2a_2^2)+
\sqrt{2\over 35}C_3a_0(6a_2^2-a_0^2)+{1\over 5}C_4(a_0^2+2a_2^2)^2+V_0 \; ,
\eeq
where $\omega$ is the rotational frequency of the nucleus and $V_0$ 
is chosen to have the minimum of the potential equal to zero. 
The dynamical variables $a_0$ and $a_2$ are connected to the deformation 
$\beta$ and asymmetry $\gamma$ by the relations 
$a_0=\beta \cos{\gamma}$ and $a_2={\beta\over \sqrt{2}}\sin{\gamma}$. 
\par
As is well known, the order--chaos transition in systems with two 
degrees of freedom may be studied by means of 
the Gaussian curvature criterion of the potential energy. 
It is, however, important to point out that {\it in general} the curvature 
criterion guarantees only a {\it local instability} and should therefore 
be combined with the Poincar\`e sections \cite{third}. 
The effective potential of the system is 
$W(a_0,a_2)= {1\over 2}B(3a_0^2+2a_2^2)\omega^2  + V(a_0,a_2)$. 
Owing to the symmetry properties of the effective potential $W$, 
our study may be restricted to the case $W(a_0,a_2=0)$. 
To apply the above criterion to our system, 
the sign of the Gaussian curvature $K$ can be obtained 
by solving the equation:
\beq
K(a_0)={\partial^2 W\over \partial a_0^2}(a_0,a_2=0)
={12\over 5}C_4 a_0^2-6\sqrt{2\over 35}C_3a_0+(C_2+3B\omega^2)=0 \; ,
\eeq
whose discriminant $\Delta$ is given by:
\beq
\Delta ={72\over 35}C_2 C_4 (\chi -{14\over 3}-14{B\omega^2\over C_2}) \; ,
\eeq
where $\chi =C_3^2/(C_2 C_4)$. 
If $\Delta \leq 0$, the curvature $K$ is always positive and the motion 
is regular. If $\Delta >0$, there is a region of negative curvature and 
the motion may be chaotic.
\par
For $C_2>0$ and $0 <\chi < 14/3$ (spherical nuclei), 
the curvature is positive and therefore the motion is regular for all 
$\omega$. For $\chi \geq 14/3$ (spherical and deformed nuclei) and 
$0\leq \omega < \sqrt{ {C_2\over 14 B}(\chi -{14\over 3})}$, 
the curvature is negative and chaotic motion may appear. 
\par
For $C_2<0$ ($\gamma$--unstable nuclei), there is a region with negative 
curvature for $0\leq \omega < \omega_c$, where 
$\omega_c=\sqrt{ {C_2\over 14 B}(\chi -{14\over 3})}$ 
is the critical frequency of the system. 
It is noteworthy that the shape of the effective potential 
$W(a_0,0)$ changes drastically as a function of $\omega$. 
If $\omega$ increases, there is a transition from chaos to order, i.e. 
the region of chaotic motion decreases and becomes zero 
for $\omega >\omega_c$. 

\section{Conclusions}

The shell model calculations show that for $Ca$ isotopes 
there are significant deviations from the Wigner distribution 
of chaotic systems. Concerning the roto--vibrational model of atomic nuclei, 
a chaos--order transition occurs as a function of the angular frequency 
of the nucleus.

\end{document}